\documentclass[useAMS,usenatbib]{mn2e}

\usepackage{amsmath}
\usepackage{lipsum} 
\usepackage{graphicx}
\usepackage{amssymb}
\usepackage{color}
\usepackage{hyperref}
\hypersetup{pdftex,  
  breaklinks=true,  
  colorlinks=true,
  urlcolor=blue,
  linkcolor=darkorange,
  citecolor=darkgreen,
  }

\definecolor{orange}{cmyk}{0,0.4,0.8,0.2}
\definecolor{darkorange}{rgb}{.71,0.21,0.01}
\definecolor{darkgreen}{rgb}{.12,.54,.11}

\newcommand\cutinhead[2]{}

\defcitealias{2014arXiv1404.4870K}{KB14}

\renewcommand\cutinhead[2]{%
 \noalign{\vskip 1.5ex}%
 \hline
\\
 \noalign{\vskip -1.5ex}%
 \multicolumn{#2}{c}{#1}%
\\
 \noalign{\vskip .8ex}%
 \hline
\\
 \noalign{\vskip -2ex}%
}%

\title[Morphology of LMC RR Lyrae stars]{Probing the distance and morphology of the Large Magellanic Cloud with RR Lyrae stars}
\author[Klein et al.]{C. R. Klein,$^{1}$\thanks{E-mail:
cklein@berkeley.edu} S. B. Cenko,$^{2}$ A. A. Miller,$^{3,4}$\thanks{Hubble Fellow.} D. J. Norman, and Joshua S. Bloom$^{1,7}$ \\
$^{1}$Astronomy Department, University of California, Berkeley, CA 94720, USA \\
$^{2}$Astrophysics Science Division, NASA Goddard Space Flight Center, Mail Code 661, Greenbelt, MD 20771, USA \\
$^{3}$Joint Space Science Institute, University of Maryland, College Park, MD 20742, USA \\
$^{4}$Jet Propulsion Laboratory, 4800 Oak Grove Drive, Pasadena, CA 91109, USA \\
$^{5}$Department of Astronomy, California Institute of Technology, Pasadena, CA 91125, USA \\
$^{6}$National Optical Astronomy Observatories, 950 N. Cherry Avenue, Tucson, AZ 85719, USA \\
$^{7}$Physics Division, 1 Cyclotron Road, Lawrence Berkeley National Laboratory, CA 94720, USA}
\begin{document}

\date{Submitted 2014 May 5.}

\pagerange{\pageref{firstpage}--\pageref{lastpage}} \pubyear{2014}

\maketitle

\label{firstpage}

\begin{abstract}

We present a Bayesian analysis of the distances to 15,040 Large Magellanic Cloud (LMC) RR Lyrae stars using $V$- and $I$-band light curves from the Optical Gravitational Lensing Experiment, in combination with new $z$-band observations from the Dark Energy Camera. Our median individual RR Lyrae distance statistical error is 1.89 kpc (fractional distance error of 3.76 per cent). We present three-dimensional contour plots of the number density of LMC RR Lyrae stars and measure a distance to the core LMC RR Lyrae centre of ${50.2482\pm0.0546~{\rm(statistical)}~\pm0.4628~{\rm(systematic)}~{\rm kpc}}$ [$\mu_{\rm LMC}=18.5056\pm0.0024~{\rm(statistical)}~\pm0.02~{\rm(systematic)}$]. This finding is statistically consistent with and four times more precise than the canonical value determined by a recent meta-analysis of 233 separate LMC distance determinations. We also measure a maximum tilt angle of $11.84^{\circ}\pm0.80^{\circ}$ at a position angle of $62^\circ$, and report highly precise constraints on the $V$, $I$, and $z$ RR Lyrae period--magnitude relations. The full dataset of observed mean-flux magnitudes, derived colour excess ${E(V-I)}$ values, and fitted RR Lyrae distances produced through this work is made available through the publication's associated online data.
\end{abstract}


\section{Introduction}
At a distance of roughly 50 kpc, the Large Magellanic Cloud (LMC) is the closest galaxy to the Milky Way with mass $\gtrsim 10^{10}~{\rm M}_\odot$. As such, it serves as an essential anchor point for the cosmic distance ladder. A precisely measured distance to the LMC allows for accurate luminosity calibrations of primary distance indicators such as pulsating variables (PVs; e.g., Cepheids and RR Lyrae stars). These calibrations enable precise distance measurements to galaxies well beyond the LMC (c.f. \citealt{2001ApJ...553...47F} and \citealt{2011ApJ...730..119R}, both of which use Cepheids to measure host galaxy distances of $>15$ Mpc), which can, in turn, calibrate secondary distance indicators such as Type Ia supernovae. \cite{2014AJ....147..122D} provides an excellent and comprehensive review of LMC distance measurements published between 1990 and the end of 2013, and recommends a canonical distance modulus of $\mu_{\rm LMC}=18.49\pm0.09$ mag (${d_{\rm LMC}=49.888\pm2.068~{\rm kpc}}$). \cite{2013Natur.495...76P} recently used eight long-period, late-type eclipsing systems in the LMC to measure a distance of $49.97\pm0.19~{\rm (statistical)}~\pm 1.11~{\rm (systematic)}~{\rm kpc}$, and this result was used in the meta-analysis as the standard by which other, less accurate findings were assessed. Precisely pinning down the LMC's distance is a necessary step in the continuing improvement of the cosmic distance ladder as progress is made towards a 1 per cent error measurement of the Hubble constant, $H_0$.

The bulk identification and study of thousands of LMC PVs also enables investigation of the three-dimensional structure of the LMC. Through their individual distance measurements, the many thousands of RR Lyrae stars in the LMC can serve as test particles tracing the old ($>10~{\rm Gyr}$) stellar population, revealing its three-dimensional morphology. While there are $\sim$10 times fewer Cepheids, these higher-mass variables can be employed to reveal the structure of the younger LMC stellar population (e.g., \citealt{1999AJ....117..920A}). With adequate precision this line of inquiry can inform our understanding of the LMC, its formation, and the nature of its gravitational interaction with the Milky Way. 

In the present study the individual distances to 15,040 RR Lyrae stars in the LMC identified in the catalogue of variable stars produced by the third phase of the Optical Gravitational Lensing Experiment (OGLE III, \citealt{2008AcA....58...89U} and \citealt{2009AcA....59....1S}) are measured and used to infer a distance to the central core of the LMC RR Lyrae population, as well as to recover the three-dimensional density structure of the population. More than a century of research has been conducted on RR Lyrae stars, much of it in pursuit of improving their utility as distance indicators, and an adequate treatment of this topic cannot be provided here. The interested reader is directed towards \citet{1964ARA&A...2...23P} and \citet{1995CAS....27.....S} which both provide excellent overviews of RR Lyrae pulsating variable stars, and towards \cite{2006ARA&A..44...93S} for a review of RR Lyrae stars as distance indicators.

This investigation is similar to the recent studies  \citep{2012AJ....144..106H,2013MNRAS.431.1565W,2014MNRAS.438.2440D}, which all used the $V$- and $I$-band OGLE III data as the basis for their LMC distance and structure measurements. Our study leverages new $z$-band mean-flux photometry for the LMC RR Lyrae stars obtained with the Dark Energy Camera (DECam, first described in \citealt{2005ASPC..339..152W} with recent status update given by \citealt{2012SPIE.8446E..11F}) in combination with the OGLE III $V$- and $I$-band light curves to simultaneously calibrate the three period--magnitude relations and fit for extinction. From these relations we determine the posterior distance moduli to each RR Lyrae star. This simultaneous period--magnitude relation calibration and distance fitting methodology was first described in \cite{2011ApJ...738..185K,2012Ap&SS.341...83K,2014MNRAS.440L..96K}, and expanded to include simultaneous colour excess fitting in \cite{2014arXiv1404.4870K}, hereafter \citetalias{2014arXiv1404.4870K}.\footnote{In the present work, however, we do not fit the colour excess as part of the model, in part because this makes the solution, now involving $10^2$ times more stars, computationally intractable.}

This paper is outlined as follows: the OGLE III and DECam data, as well as the calculation of mean-flux magnitudes, are described in Section\,\ref{LMC_data_desc}. The method by which the individual RR Lyrae star distances are determined is detailed in Section\,\ref{RRL_distances}. In Section\,\ref{LMC_distance} these distances are used to describe and analyse the distance to and morphology of the LMC RR Lyrae population. Finally, in Section\,\ref{conclusions} we compare the present work to other similar studies and discuss the implications of our results.

\section[]{Data Description}\label{LMC_data_desc}

The OGLE III catalogue of LMC RR Lyrae stars (overview of LMC photometry: \citealt{2008AcA....58...89U}, characterisation of LMC RR Lyrae stars: \citealt{2009AcA....59....1S}) was used as the input target list for the present work. This is supplemented with $z$-band observations of the LMC acquired with DECam during its science verification period in 2012 November. The OGLE III catalogue provided $V$- and $I$-band light curves for 22,247 stars. Requiring that each star also have DECam $z$-band data results in a sample of 17,629 stars. This is further reduced by median absolute deviation clipping and sigma-clipping from a least squares linear regression, described in detail in subection\,\ref{LMC_pmrs}, to a final sample of 15,040 RR Lyrae stars (11,846 RRab and 3,194 RRc).

\begin{figure}
	\centering
	\includegraphics{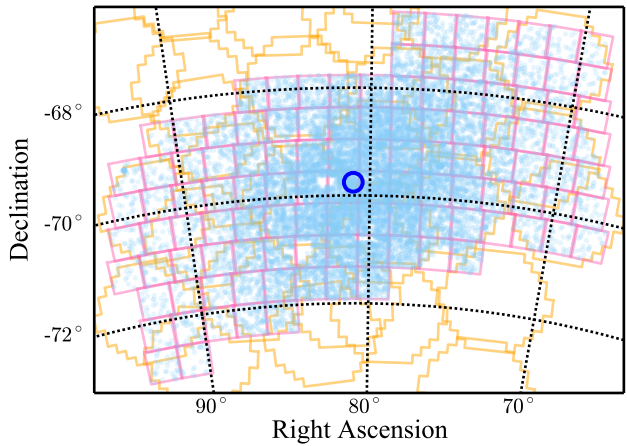}
	\caption{Map of the LMC RR Lyrae stars (blue points) superimposed upon OGLE III (pink) and DECam (gold) field-of-view outlines. The central blue circle shows the optical centre of the LMC at RA: $80.8942^{\circ}$, Dec: $-69.7561^{\circ}$ (J2000).}
	\label{fig:lmc_projected_density}
\end{figure}

The sky coverage of both the OGLE III survey and the DECam observing program, along with the scatter plot map of the final RR Lyrae sample, is shown in Fig.\,\ref{fig:lmc_projected_density}. The OGLE III coverage is quite uniform and complete within its combined footprint, but the DECam coverage has multiple gaps resulting from the observing strategy and instrumental deficiencies which are further discussed in subsection \ref{decam_data}.

\subsection[]{OGLE III mean-flux magnitudes}\label{decam_data}

OGLE III $V$- and $I$-band light curves and measured periods for the RR Lyrae stars (RRab and RRc subtypes) were downloaded from the catalogue published in \cite{2009AcA....59....1S}. Mean-flux magnitudes for each light curve were measured following the procedure described in Section 3 of \citetalias{2014arXiv1404.4870K}. In short, the raw data is parametrically resampled 500 times to fit 500 harmonic models and derive 500 mean-flux magnitude measurements for each light curve. The standard deviation of these bootstrapped mean-flux magnitude measurements is taken to be the measurement error. This procedure resulted in 22,125 $V$-band and 22,188 $I$-band RR Lyrae mean-flux magnitudes. This is less than the starting OGLE III dataset of 22,247 light curves in each band because light curves which produce highly discrepant bootstrapped harmonic models are rejected. The mean error in $V$ is 0.006 mag and the mean error in $I$ is 0.0024 mag.

\subsection[]{DECam observations and reduction}\label{decam_data}

As part of the DECam Science Verification program (Program ID: 2012B-3002; 
PI: Bloom), we obtained a single epoch of $z$-band exposures of 30 $\times$
3.0\,deg$^{2}$ fields on 16 nights from 2012 November 1--18 UT.  The fields
were selected to maximise coverage of the OGLE III LMC footprint, with
small offsets applied to cover chip gaps and minimise bleed trails from
extremely bright stars (Fig.\,\ref{fig:lmc_projected_density}).  For each
exposure, 61 of the 62 individual science CCDs\footnote{One chip, C61, was
not fully operable during the Science Verification run.} were processed 
using standard reduction algorithms (bias subtraction, flat-fielding, etc.) 
using the computational resources at the National Energy Research Scientific 
Computing Center (NERSC\footnote{See \url{http://www.nersc.gov}.}).  

Astrometry on individual frames was calibrated with respect to reference
sources from the Two Micron All Sky Survey (2MASS; \citealt{2006AJ....131.1163S}) using
the astrometry.net software package \citep{lhm+10}.  Photometric calibration
was performed using same-night observations of sources in the Sloan Digital
Sky Survey Stripe 82 Standard Star Catalogue \citep{ism+07}.  Applying
standard calibration methodology (e.g., \citealt{ols+12}), we find
that we can achieve a robust scatter in our absolute photometric calibration
of $\lesssim 0.02$\,mag on clear nights ($\gtrsim 50$ per cent of the observing
time from the Science Verification run).  While this calibration can be 
improved with more advanced modeling of instrumental signatures (e.g.,
\citealt{taa+14}), we find that 2 per cent precision is sufficient for our
scientific objectives.\footnote{Error in the colour excess, itself dominated by intrinsic scatter in the previously-derived period--magnitude relations of \citetalias{2014arXiv1404.4870K}, dominates over photometry errors when calculating individual RR Lyrae star distances.}

The DECam observing program produced on average one 1-second exposure for each of the 30 fields each night. This sub-optimal exposure depth and cadence was necessitated by the oversubscription of DECam and the desire to test the instrument performance in unusual or extreme modes of operation during the science verification period. Most of the RR Lyrae targets are marginally detected in single exposures, but this was not sufficient to produce the traditional phase-folded light curves that provide mean-flux measurements through harmonic modeling. To recover $z$-band mean-flux magnitudes the individual epochs of each CCD were flux-scaled using relative photometric zero points measured with PSFex \citep{2011ASPC..442..435B} and SExtractor \citep{1996A&AS..117..393B}, and then average combined with Swarp \citep{2002ASPC..281..228B}. This procedure resulted in a mean error on the $z$-band mean-flux magnitude measurements for the final RR Lyrae sample of 0.0387 mag, which includes the errors introduced by absolute photometric calibration and the relative epoch-to-epoch flux-scaling.

\section[]{RR Lyrae Distance Measurements}\label{RRL_distances}

Distances for the individual RR Lyrae stars are measured using the observed, extinction-corrected $V$, $I$, and $z$ magnitudes in combination with the period--magnitude relations. The method employed is similar to the simultaneous Bayesian linear regression methodology described in \citetalias{2014arXiv1404.4870K}. A significant difference is that in the present analysis the colour excess is considered part of the observed data, not as a prior to which a posterior distribution is fit. The following two subsections detail the derivation of individual RR Lyrae colour excess and provide more description of the specific period--magnitude relations fitting procedure.

\subsection[]{Colour excess}\label{colour_excess_sec}

The ${E(V-I)}$ colour excess for each RR Lyrae star is derived from the observed OGLE III mean-flux magnitudes and the previously-calibrated $V$ and $I$ period--magnitude relations published in \citetalias{2014arXiv1404.4870K}. This approach is conceptually similar to that of \cite{2011AJ....141..158H}, with the main difference being that the earlier study used the theoretical $V$-band metallicity--luminosity and $I$-band period--metallicity--luminosity relations of \cite{2004ApJS..154..633C}. In the present work, colour excess is given by the subtraction of the absolute colour (from the period--magnitude relations) from the observed colour,
\begin{equation}
\label{eqn:colour_excess}
E(V-I) = (m_V - m_I) - [M_V(P) - M_I(P)].
\end{equation}

\begin{figure}
	\centering
	\includegraphics{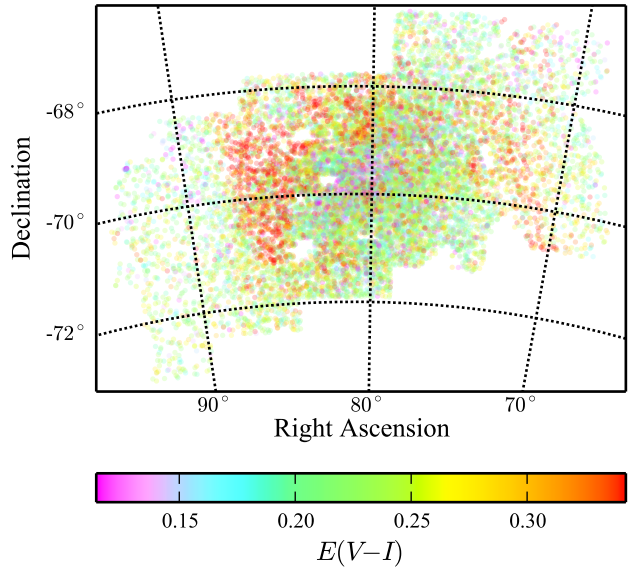}
	\caption{Map of the LMC RR Lyrae stars coloured by colour excess value, ${E(V-I)}$. Median RR Lyrae colour excess value is 0.228 mag, and the median error per star is 0.093 mag.}
	\label{fig:lmc_projected_EVI}
\end{figure}

The dominant source of error in the colour excess calculation is the intrinsic scatter of the period--magnitude relations. The median colour excess for the LMC RR Lyrae population is found to be 0.228 mag, with a median error of 0.093 mag. This is significantly greater than the median value of 0.11 mag (with standard deviation of 0.06 mag) found by \cite{2011AJ....141..158H}. Fig.\,\ref{fig:lmc_projected_EVI} is a map of the RR Lyrae distribution coloured by color excess. Two prominent regions of large extinction are apparent, one shaped like a downward-pointing wedge located at a right ascension $\approx87^\circ$, and the other a band running north-south centred at right ascension $\approx73^\circ$. Both of these features are also noted by \cite{2011AJ....141..158H} and depicted in their Fig.\,10. 

The band-specific extinction for each star was derived from the measured colour excess value using the extinction curve data given in Table\,6 of \cite{1998ApJ...500..525S}. We apply a conversion factor of 1.62 to transform ${E(V-I)}$ to the conventional ${E(B-V)}$ (see \citealt{1968nim..book..167J}, \citealt{1975A&A....43..133S}, and \citealt{1985ApJ...288..618R}). The corrected mean-flux magnitudes are thus given by
\begin{eqnarray}
m_V &=& m_{V,{\rm obs}} - 3.240 \times \left[E\left(V-I\right)/1.62\right] \label{eqn:corrected_V} \\ 
m_I &=& m_{I,{\rm obs}} - 1.962 \times \left[E\left(V-I\right)/1.62\right] \label{eqn:corrected_I} \\ 
m_z &=& m_{z,{\rm obs}} - 1.479 \times \left[E\left(V-I\right)/1.62\right]. \label{eqn:corrected_z}
\end{eqnarray}

\subsection[]{Period--magnitude relations}\label{LMC_pmrs}

The $V$, $I$, and $z$ extinction-corrected mean-flux magnitudes were used to calibrate period--magnitude relations through a method similar to the Bayesian simultaneous linear regression formalism employed for 13 simultaneous fits in \citetalias{2014arXiv1404.4870K}. The primary difference in this application is that the colour excess is not fitted as a model parameter, and is instead incorporated into the likelihood (observed data). The framework easily accommodates the extra model parameters, but the augmented processing time, which goes roughly as $\mathcal{O}(n^2)$, is unreasonable for fitting a model with 15,040 stars (compared to the calibration sample size of 134 for \citetalias{2014arXiv1404.4870K}).

Before the Bayesian MCMC fitting procedure was performed, the dataset of 17,629 stars was cleaned to reject outliers. These are most likely foreground stars or stars with poorly measured photometry resulting from crowding effects. All stars with a median absolute deviation in magnitude greater than $5\sigma$ for any of the three wavebands were removed, and then a simple least squares linear regression was performed to fit preliminary period--magnitude relations and all stars more than $4\sigma$ from the best fitted line for any waveband's relation were also removed. 15,040 RR Lyrae stars survived the cuts and made it into the calibration sample.

The calibration sample is composed of 11,846 RRab stars (fundamental mode pulsators, period $P_{f}$) and 3,194 RRc stars (first overtone pulsators, periods $P_{fo}$). The RRc stars' periods must be ``fundamentalised'' before deriving the period--magnitude relations. As in \cite{2004ApJ...610..269D}, an RRc star's fundamentalised period is given by 
\begin{equation}
\label{eqn:period_fundamentalize}
\log_{10}\left(P_{f}\right) = \log_{10}\left(P_{fo}\right) + 0.127.
\end{equation}
The general form of the period--magnitude relation is then 
\begin{equation}
\label{eqn:general_PMR}
m_{ij}=\mu_i + M_{0,j} + \alpha_j \log_{10}\left(P_i/P_0\right) + \epsilon_{ij},
\end{equation}
where $m_{ij}$ is the observed apparent, extinction-corrected mean-flux magnitude of the $i$th RR Lyrae star in the $j$th waveband, $\mu_i$ is the distance modulus for the $i$th RR Lyrae star, $M_{0,j}$ is the absolute magnitude zero point for the $j$th waveband, $\alpha_j$ is the slope in the $j$th waveband, $P_i$ is the fundamentalised period of the $i$th RR Lyrae star in days, $P_0$ is a period normalisation factor (for consistency with \citetalias{2014arXiv1404.4870K} we use ${P_0 = 0.52854~{\rm d}}$), and the $\epsilon_{ij}$ error terms are independent zero-mean Gaussian random deviates with variance ${(\sigma_{{\rm intrinsic},j}^2 +  \sigma_{m_{ij}}^2)}$.

The error on the extinction-corrected mean-flux magnitudes, $\sigma_{m_{ij}}$, was derived by propagating the error from the contributing observed apparent magnitudes and colour excess terms (see equations\,\ref{eqn:corrected_V}--\ref{eqn:corrected_z}). The intrinsic scatter of the period--magnitude relations, $\sigma_{{\rm intrinsic},j}$, which is added in quadrature with $\sigma_{m_{ij}}^2$ to calculate the standard deviation of the likelihood, is adopted from the findings of \citetalias{2014arXiv1404.4870K}: ${\sigma_{{\rm intrinsic},V} = 0.0320}$, ${\sigma_{{\rm intrinsic},I} = 0.0713}$, and ${\sigma_{{\rm intrinsic},z} = 0.1153}$.

The prior distributions for $M_{0,j}$ and $\alpha_j$ were normal distributions centred at the fitted values for the $V$, $I$, and $z$ period--magnitude relations found by \citetalias{2014arXiv1404.4870K}, with standard deviations expanded to 0.2 for $M_0$ and 1.5 for $\alpha$ (to allow the MCMC traces freedom to explore a wider parameter-space). The same prior, $\mathcal{N}(18.5, 0.2163^2)$, was used for all of the $\mu_i$. This standard deviation was selected to be a fractional distance error of 10 per cent (${\approx 5~{\rm kpc}}$), which is much larger than the depth of the LMC and significantly larger than ($>2$ times) the median posterior $\sigma_{\mu_i}$.

To fit the model given by equation\,\ref{eqn:general_PMR} ten identical MCMC traces were run, each generating 3.5 million iterations. The first 0.5 million were discarded as burn-in and the remaining 3 million were thinned by 300 to result in ten traces of 10,000 iterations each. The Gelman-Rubin convergence diagnostic, $\hat{R}$ \citep{gelman_rubin_convergence}, was computed for each posterior model parameter (3 zero points, 3 slopes, and 15,040 distance moduli) and all are found to be well-converged (${\hat{R} < 1.1}$). 

\begin{figure}
	\centering
	\includegraphics{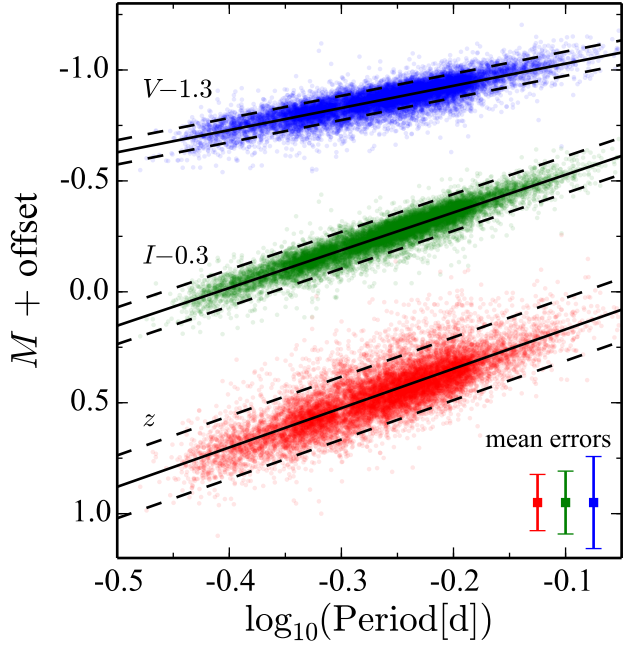}
	\caption{$V$-, $I$-, and $z$-band period--magnitude relations (solid lines) derived for the LMC RR Lyrae population, superimposed on scatter plots of the RR Lyrae posteriors ($M$ computed using $\mu_{\rm Post}$). The dashed lines denote the $1\sigma$ prediction intervals for a new RR Lyrae star with known period.}
	\label{fig:pmrs}
\end{figure}

The best fitted period--magnitude relations and a scatter plot of the RR Lyrae posteriors ($M$ computed using $\mu_{\rm Post}$) is presented in Fig.\,\ref{fig:pmrs}. The equations for the period--magnitude relations are 
\begin{eqnarray}
M_V=\left(0.448 \pm 0.003\right) - \left(0.999 \pm 0.038\right) \times \log_{10}\left(P/P_0\right) \label{eqn:V_pmr} \\ 
M_I=\left(0.073 \pm 0.002\right) - \left(1.701 \pm 0.034\right) \times \log_{10}\left(P/P_0\right) \label{eqn:I_pmr} \\ 
M_z=\left(0.483 \pm 0.002\right) - \left(1.774 \pm 0.034\right) \times \log_{10}\left(P/P_0\right) . \label{eqn:z_pmr}
\end{eqnarray}
These results are consistent (within $2\sigma$) with the findings published in \citetalias{2014arXiv1404.4870K}. The new slopes are systematically lower, although the previous constraints are considerably wider. The extremely tight distributions for the posterior $M_0$ and $\alpha$ are due to the very large number of RR Lyrae stars in the calibration dataset, as compared to previous studies that have used calibration samples of a few dozen to slightly more than one hundred stars collated from the local Milky Way field RR Lyrae population.

\section[]{LMC Distance and Morphology}\label{LMC_distance}

\begin{figure}
	\centering
	\includegraphics{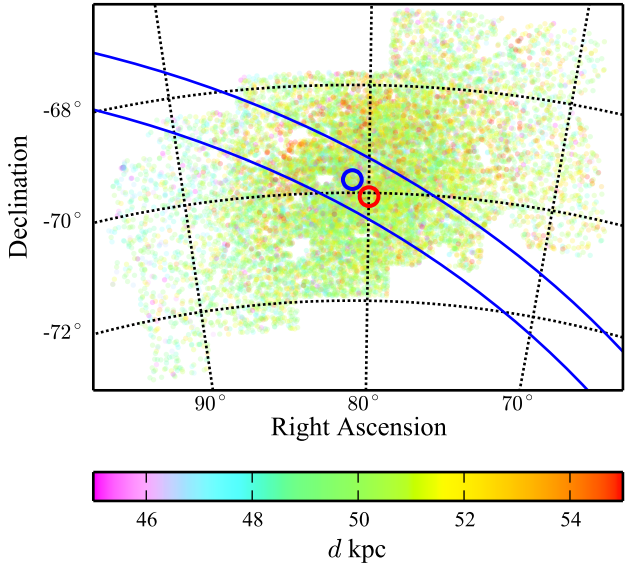}
	\caption{Map of the LMC RR Lyrae stars coloured by distance. The strip outlined in blue is centred at the LMC optical centre (blue circle), and has a width of ${1~{\rm kpc}}$ and a position angle of $62^{\circ}$. Along this strip the distance slope angle (the tilt of the LMC in the plane of the sky) is measured to be $11.84^{\circ}\pm0.80^{\circ}$, with the eastern side (left-hand-side on the page) being closer to Earth. The red circle denotes our measured central position for the core RR Lyrae population at right ascension $79.9855^\circ$ and declination $-70.0697^\circ$.}
	\label{fig:lmc_distance_map}
\end{figure}

The individual distance moduli fitted via the Bayesian simultaneous linear regression method described in subsection\,\ref{LMC_pmrs} have a median error of 0.0816 mag (a fractional distance error of 3.76 per cent, or ${1.89~{\rm kpc}}$). The standard deviation of the distances is 2.2 kpc, which is a proxy for the physical extent of the LMC along the line of sight. Thus, the individual RR Lyrae distances serve as a probe of the LMC depth structure, and can be analysed as a population to reveal structure at even smaller physical scales. The following subsections present an analysis of the spatial distribution of the LMC core RR Lyrae population and the overall tilt of the LMC in the plane of the sky. Fig.\,\ref{fig:lmc_distance_map} shows a map of the RR Lyrae stars coloured by distance, with the strip exhibiting maximum tilt outlined in blue.

\subsection[]{RR Lyrae density structure}\label{RRL_density}

\begin{figure}
	\centering
	\includegraphics{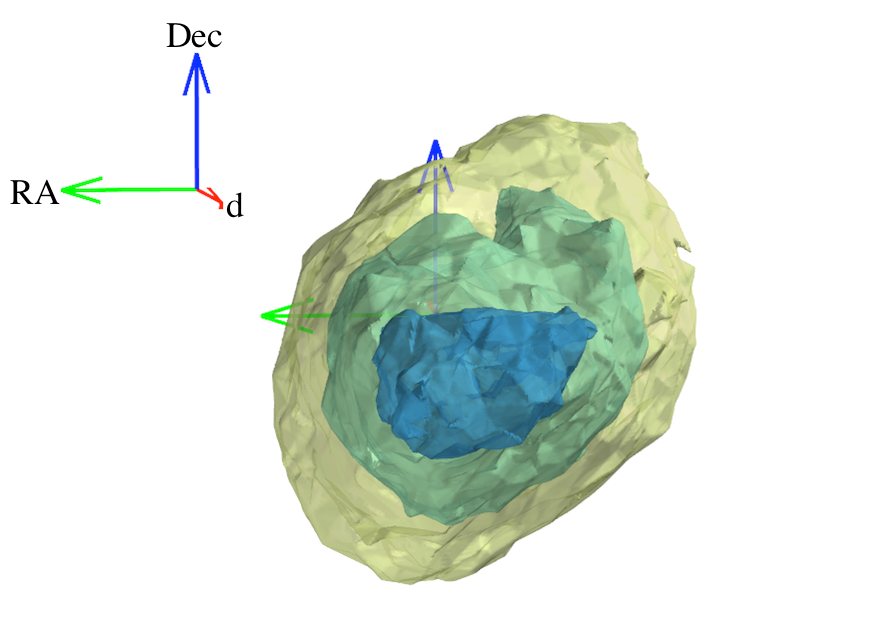}
	\caption{Three-dimensional contour plot of RR Lyrae number density in the core of the LMC. The view is projected along the vector pointing away from Earth (red arrow pointing into the page). The green arrow points along the direction of increasing right ascension and the blue arrow points along the direction of increasing declination. The origin of the arrow vectors inside the density contours is at the optical centre of the LMC, at the central distance of the RR Lyrae population. Each arrow is 1 kpc in length. The contour surfaces are at RR Lyrae number densities of 200, 250, and ${300~{\rm kpc}^{-3}}$.}
	\label{fig:threeD_R}
\end{figure}

\begin{figure}
	\centering
	\includegraphics{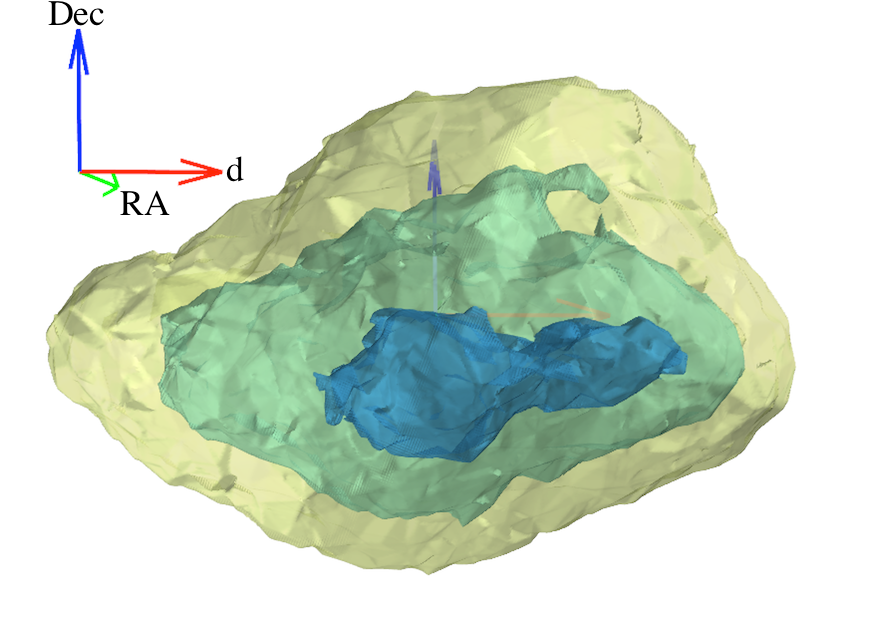}
	\caption{Three-dimensional contour plot of RR Lyrae number density in the core of the LMC, now rotated so that the view is projected along the vector of increasing right ascension (right ascension increases into the page). Axis arrows and contours same as Fig.\,\ref{fig:threeD_R}. }
	\label{fig:threeD_RA}
\end{figure}

\begin{figure}
	\centering
	\includegraphics{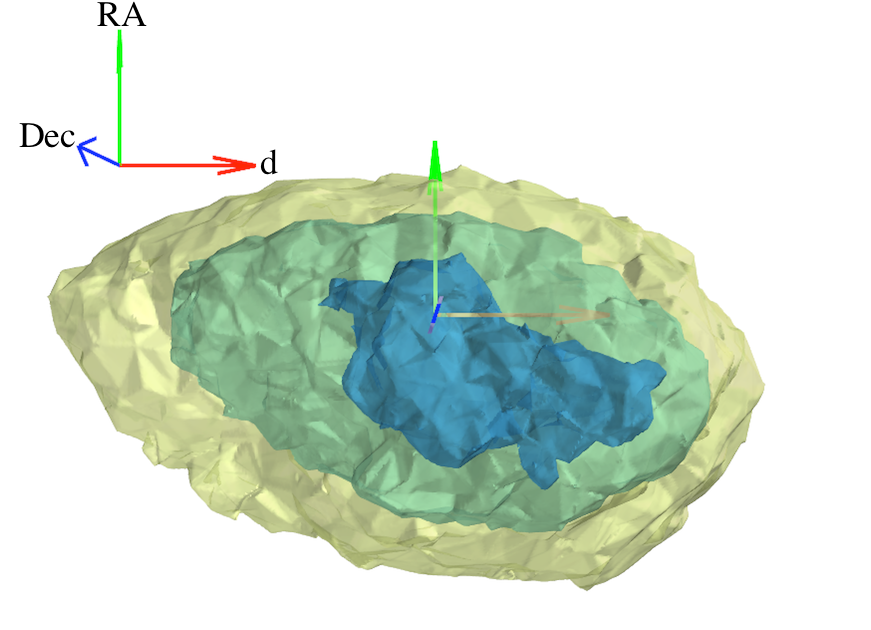}
	\caption{Three-dimensional contour plot of RR Lyrae number density in the core of the LMC, now rotated so that the view is projected along the vector of decreasing declination (declination increases out of the page). Axis arrows and contours same as Fig.\,\ref{fig:threeD_R}. This view illustrates the tilt of the LMC in the plane of the sky. We see here that eastern side (larger right ascension) is generally closer to Earth.}
	\label{fig:threeD_Dec}
\end{figure}

To investigate the density structure of RR Lyrae stars in the LMC, the spherical sky coordinates (right ascension, declination, and distance from Earth) for each star were transformed into a local, LMC-centred cartesian coordinate frame. Then, the local number density was computed for each star by counting the number of neighboring stars within a sphere of $V=1~{\rm kpc}^3$. This generated 15,040 local number density data points which were then interpolated onto a grid of $401\times401\times401$ voxels in a cube of side length 6 kpc to produce the three-dimensional contour plots presented in Figs.\,\ref{fig:threeD_R}, \ref{fig:threeD_RA}, and \ref{fig:threeD_Dec}.

These three-dimensional plots clearly show that the core, highest-density concentration of RR Lyrae stars lies southward and somewhat westward of the optical centre. Additionally, the depth of the core appears  significantly larger than its extent in right ascension or declination. The much larger individual RR Lyrae position error in depth ($\sim1.9~{\rm kpc}$ vs effectively 0 for right ascension and declination) can lead to apparent elongation along that axis in these results, which must be taken into account when interpreting the plots.

A distance to the centre of the core LMC RR Lyrae population was determined by parametric resampling of the mean distance measurement for the 1,231 stars that lie within the $250~{\rm kpc}^{-3}$ density contour. This resulting distance measurement is ${50.2482\pm0.0546~{\rm kpc}}$ ($\mu_{\rm LMC}=18.5056\pm0.0024$), where the given error is statistical. Due to the 0.02 mag absolute photometric calibration of the DECam $z$-band mean-flux magnitudes [which dominates over the OGLE III photometric calibration error \citep{2008AcA....58...69U}], an additional systematic LMC distance error of 0.4628 kpc (0.02 mag for distance modulus) is appropriate. The right ascension of the core centre was found to be $79.9855^\circ$ and the declination was found to be $-70.0697^\circ$.

\subsection[]{Tilt of the LMC RR Lyrae population}\label{RRL_tilt}

\begin{figure}
	\centering
	\includegraphics{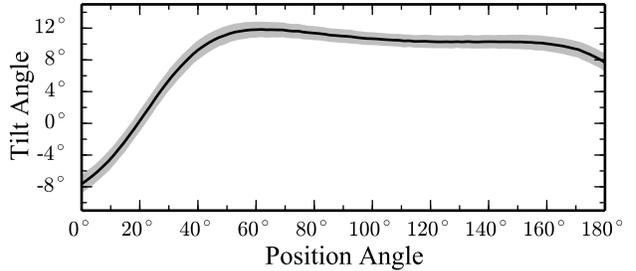}
	\caption{LMC tilt angle measured along 1 kpc wide strips centered at the LMC optical centre. The gray shaded band is the standard deviation found by parametrically resampling the RR Lyrae distance data 1000 times. Maximum tilt angle of $11.84^{\circ}\pm0.80^{\circ}$ is measured at a position angle of $62^\circ$. This strip is superimposed on the LMC RR Lyrae map shown in Fig.\,\ref{fig:lmc_distance_map}.}
	\label{fig:LMC_tilt}
\end{figure}

The tilt angle of the LMC in the plane of the sky (derived from the depth slope) was measured by rotating 1 kpc-wide strips about the optical centre through all position angles ($0^\circ$ through $180^\circ$). The general schematic for this approach is provided in Fig.\,\ref{fig:lmc_distance_map}. All of the stars within a strip were projected onto the strip centre line to provide a consistent metric for distance along the strip, akin to radial distance from the LMC centre. Then, 1000 least squares linear regressions were performed with parametric resampling to measure the slope of the mean LMC depth along each strip. This procedure was conducted at 100 position angles to produce the results shown in Fig.\,\ref{fig:LMC_tilt}. The maximum LMC tilt angle of $11.84^{\circ}\pm0.80^{\circ}$ is measured at a position angle of $62^\circ$. It is also important to note that a wide plateau of tilt angle $>8^\circ$ is observed between position angles of $40^\circ$ and $180^\circ$, quite consistent with an overall LMC position angle $\approx 110^\circ$.

The tilt angle found through this method is in significant disagreement with the much larger inclination angle values found by recent studies, $24.20^\circ$ (no error given) and $32^\circ\pm4^\circ$ by \citealt{2014MNRAS.438.2440D} and \citealt{2012AJ....144..106H}, respectively. The full dataset of RR Lyrae distances found through the present work is available in the publication's associated online data, and we encourage interested researchers to apply their favored spatial modeling technique to our data to corroborate or disprove the results of this analysis.

\section[]{Discussion and Conclusions}\label{conclusions}

We have combined the OGLE III $V$- and $I$-band LMC RR Lyrae light curve data with new $z$-band observations from DECam to measure the distance to 15,040 LMC RR Lyrae stars and simultaneously fit the $V$, $I$, and $z$ period--magnitude relations. Our primary findings are much tighter constraints on the period--magnitude relation zero points and slopes, as well as a new, precise distance measurement to the centre of the core LMC RR Lyrae population of ${50.2482\pm0.0546~{\rm(statistical)}~\pm0.4628~{\rm(systematic)}~{\rm kpc}}$ [$\mu_{\rm LMC}=18.5056\pm0.0024~{\rm(statistical)}~\pm0.02~{\rm(systematic)}$]. This finding is statistically consistent with and four times more precise than the canonical value determined by \cite{2014AJ....147..122D} through a meta-analysis of 233 separate LMC distance determinations published between 1990 and 2013.

We additionally provide three-dimensional contour plots of the RR Lyrae number density distribution in Figs.\,\ref{fig:threeD_R}, \ref{fig:threeD_RA}, and \ref{fig:threeD_Dec} which aide in visualising the location of the centre of the RR Lyrae population with respect to the LMC optical centre and the tilt of the LMC in the plane of the sky (particularly apparent in Fig.\,\ref{fig:threeD_Dec}). We conducted an analysis to measure the tilt angle of the LMC in the plane of the sky and found the maximum tilt to be $11.84^{\circ}\pm0.80^{\circ}$ at a position angle of $62^\circ$.

The full dataset of 15,040 RR Lyrae stars with mean-flux magnitude measurements in $V$, $I$, and $z$, along with derived colour excess ${E(V-I)}$ values and fitted distances are provided in this publication's associated online data. We encourage other researchers to conduct independent analyses of LMC RR Lyrae morphology and overall distance using our dataset.

We caution against future studies that use a singular, highly precise LMC distance measurement as the basis for calibrating distance indicators. While the centre point of the LMC can be well-defined by distances to a few thousand stars, the uncertainty to any one star is still about 3-4 per cent. Thus the distance to individual calibrators (i.e., Cepheids) in the LMC system cannot be determined as precisely as our reported distance to the core LMC RR Lyrae centre. At worst, the distance to any single member of the LMC can be inferred with error equal to the spread in the LMC depth, which we find to be 2.22 kpc. 

The best way to improve RR Lyrae-based LMC distance measurements is to incorporate longer-wavelength photometry, either ground-based near-infrared or space-based mid-infrared (the latter being even more beneficial). This will enable the use of the higher-precision period--magnitude relations applicable for $\lambda > 1~\mu{\rm m}$ derived in \citetalias{2014arXiv1404.4870K}. Additionally, these data would require lower extinction correction values, further reducing the error on the RR Lyrae absolute magnitudes. Finally, longer-wavelength and multi-waveband data, in combination with continued development of the MCMC sampling algorithms and augmented computational resources, can allow for simultaneous fitting of absolute magnitude and colour excess for the tens of thousands of LMC RR Lyrae stars.

\section*{Acknowledgments}

Authors Klein and Bloom acknowledge the generous support of grants (\#0941742, \#1009991, \#1251274) from the National Science Foundation. 
A.A.M. acknowledges support for this work by NASA from a Hubble Fellowship grant HST-HF-51325.01, awarded by STScI, operated by AURA, Inc., for NASA, under contract NAS 5-26555. Part of the research was carried out at the Jet Propulsion Laboratory, California Institute of Technology, under a contract with the National Aeronautics and Space Administration. 
The authors thank Rollin Thomas and Peter Nugent for developing and adapting the DECam data processing systems used in the course of this work.
This project used data obtained with the Dark Energy Camera (DECam), which was constructed by the Dark Energy Survey (DES) collaborating institutions: Argonne National Lab, University of California Santa Cruz, University of Cambridge, Centro de Investigaciones Energeticas, Medioambientales y Tecnologicas-Madrid, University of Chicago, University College London, DES-Brazil consortium, University of Edinburgh, ETH-Zurich, University of Illinois at Urbana-Champaign, Institut de Ciencies de l`Espai, Institut de Fisica d`Altes Energies, Lawrence Berkeley National Lab, Ludwig-Maximilians Universitat, University of Michigan, National Optical Astronomy Observatory, University of Nottingham, Ohio State University, University of Pennsylvania, University of Portsmouth, SLAC National Lab, Stanford University, University of Sussex, and Texas A\&M University. Funding for DES, including DECam, has been provided by the U.S. Department of Energy, National Science Foundation, Ministry of Education and Science (Spain), Science and Technology Facilities Council (UK), Higher Education Funding Council (England), National Center for Supercomputing Applications, Kavli Institute for Cosmological Physics, Financiadora de Estudos e Projetos, Funda\c{}\~ao Carlos Chagas Filho de Amparo a Pesquisa, Conselho Nacional de Desenvolvimento Cient\'{\i}fico e Tecnol\'{—}gico and the Minist\'erio da Ci\^encia e Tecnologia (Brazil), the German Research Foundation-sponsored cluster of excellence ``Origin and Structure of the Universe'' and the DES collaborating institutions.
The National Energy Research Scientific Computing Center, supported by the Office of Science of the U.S. Department of Energy, provided staff, computational resources, and data storage for this project.
This research has made use of the SIMBAD database, operated at CDS, Strasbourg, France.
This research has made use of NASA's Astrophysics Data System. 

\bibliographystyle{mn2e_alt}
\bibliography{Klein_refs}


\label{lastpage}

\end{document}